\documentclass{ws-procs9x6}

\newcommand{\beq}{\begin{equation}}
\newcommand{\eeq}{\end{equation}}
\newcommand{\beqa}{\begin{eqnarray}}
\newcommand{\eeqa}{\end{eqnarray}}
\newcommand{\om}{\Omega_m}

\begin{document}

\title{Probing dark energy with SNAP}

\author{Eric Linder}

\address{Physics Division, Berkeley Lab \\ 
Berkeley, CA 94720, USA\\ 
E-mail: evlinder@lbl.gov}

\maketitle

\abstracts{The distance-redshift relation observed for 
supernovae has led to the 
discovery that the expansion of the universe is accelerating.  A next 
generation experiment, the Supernova/Acceleration Probe (SNAP), can 
investigate the nature of the dark energy responsible, determining 
its energy density contribution and equation of state.  In 
addition, indications of time variation in the equation of state could 
provide critical clues to the underlying fundamental physics. 
}

\section{Dark Energy Today}

The discovery in 1998 \cite{perl,hiz} that the expansion of the universe 
is accelerating energized both the cosmology and particle physics 
communities.  Whatever the physical mechanism -- 
a cosmological constant, dynamical scalar field, higher dimensions, 
or altered gravitation -- it lies beyond the current standard model 
of physics.  Moreover, such an acceleration requires an unknown 
component to contribute the majority of the energy density of the universe -- 
twice as much as the matter density.  Researchers are intently 
designing theories to explain these observations as well as experiments 
to test the theories and probe the nature of this ``dark energy''. 

Type Ia supernovae observations have a clear, simple connection to the 
expansion of the universe that makes them excellent at probing the 
dynamics out to redshifts $z\approx2$.  The apparent magnitude, or 
intensity, of the supernova is a measure of the distance and its redshift 
directly translates into the expansion factor, mapping out the recent 
expansion history of the universe. 

Other cosmological probes exist as well, and early data from them have 
confirmed the original supernova results.  From the cosmic microwave 
background (CMB), the locations of the acoustic peaks in the temperature 
power spectrum coincidentally lie along contours of the age of the 
universe \cite{knox}, with their measurement implying 
a total age $14.0\pm0.5$ billion years.  Such a high number requires a 
substantial negative equation of state component in addition to matter, 
at least for conventional values of the Hubble constant.  Note though 
that this argument does rely on the universe being flat and the 
matter density perturbations adiabatic.  The detailed structure of 
the CMB power spectrum gives further information: currently it is best 
fit with a cosmological constant contribution to the dimensionless energy 
density of $\Omega_\Lambda=0.53-0.70$ \cite{bond}.  This as well depends 
on subsidiary knowledge, e.g.~neglecting the gravitational wave contribution 
to the power. 

From large scale galaxy redshift surveys such as 2dF, the  
power spectrum of matter density fluctuations also points to an 
accelerating universe through its shape \cite{teghamxu}.  
The ``look'' of large 
scale structure in computer simulations agrees as well with the 
presence of a large cosmological constant like component.  In the 
next few years measurements of the growth rate of evolving structure 
may become precise enough to give further evidence.  Combining CMB 
and large scale structure provides strong evidence for a universe 
consisting 1/3 of matter and 2/3 of some dark energy \cite{bond,efst}; 
this is independent of and fully consistent with the supernova 
measurements.  So today we have a concordance that the universe 
is accelerating, its energy dominated by dark energy with a 
strongly negative equation of state. 

\section{The Next Generation} 

But we know almost nothing of the nature of the dark energy -- its 
equation of state ratio of pressure to density, $w=p/\rho$, or whether 
this evolves, e.g.~$w'$.  These two quantities hold crucial 
clues to the underlying fundamental physics.  By mapping the expansion 
history of the universe we can investigate the properties of the new 
physics.  Illustratively, distance-redshift observations $d(z)$ teach 
us the cosmological dynamics, which in turn guides us to the high energy 
field theory: 
\beq 
d(z) \quad\longrightarrow\quad a(t) \quad\longrightarrow\quad 
V(\phi(a(t))) 
\eeq 
where $V(\phi)$ is the potential of the dark energy field. 

A next generation experiment needs 
to be carefully designed to probe the dark energy.  But systematic 
uncertainties rather than merely paucity or imprecision of observations 
will be the key obstacle.  The CMB is insensitive to time 
variation and can only provide a rough estimate of an averaged value of 
$w$, except for a small impact on the late time Sachs-Wolfe effect buried 
in cosmic variance.  Gravitaional lensing and the growth rate of large 
scale structure is promising, 
but needs to be separated from complicated nonlinear astrophysics.  
Supernovae studies directly measure the expansion dynamics and 
have a longer history than the various structure related methods.  
From this history a comprehensive list of possible systematics, 
and methods for accounting for them, have been developed and a new 
experiment is under design dedicated to exploring dark energy: the 
Supernova/Acceleration Probe (SNAP) satellite \cite{snap}. 

To investigate the dark energy and distinguish between classes of 
physics models we need to probe the expansion back into the 
matter dominated 
deceleration epoch, indeed over a redshift baseline reaching $z>1.5$ 
\cite{lh02}.  The SNAP mission concept is a 2.0 meter space telescope 
with a nearly 
one square degree field of view.  A half 
billion pixel, wide field imaging 
system comprises 36 large format new technology CCD's and 36 
HgCdTe infrared detectors.  Both the imager and a low resolution ($R\sim 100$) 
spectrograph cover the wavelength range 3500 - 17000 \AA, allowing 
detailed characterization of Type Ia supernovae out to $z=1.7$. 

As a space experiment SNAP will be able to study supernovae over a much 
larger range of redshifts than has been possible with the current 
ground-based measurements -- over a wide wavelength range unhindered by 
the Earth's atmosphere and with much higher precision and accuracy. 

An array of data (e.g.~supernova 
risetime, early detection to eliminate Malmquist bias, 
lightcurve peak-to-tail ratio, 
spectrum, separation of supernova light from host galaxy 
light, identification of host galaxy morphology, etc.) 
makes it possible to study each individual supernova and measure 
enough of its physical properties to recognize deviations from standard 
brightness subtypes.  
For example, an approach to the problem of possible supernova evolution 
\cite{coping} uses the rich stream of information that an expanding supernova 
atmosphere sends us in the form of its spectrum.  
A series of measurements will be constructed for each supernova that 
define systematics-bounding subsets of the Type Ia category.  
Only the change in brightness as a function of the 
parameters classifying a subtype is needed, not any intrinsic brightness. 
By matching like to like among the supernova subtypes over redshift, we can 
construct independent distance-redshift, or Hubble, diagrams for each, 
which when compared bound 
systematic uncertainties at the targeted level of 0.02 magnitudes.

\section{Revealing the Nature of Dark Energy} 

Generically the equation of state of dark energy is not constant in 
time and indeed its evolution provides critical information on the 
class of physics responsible for the accelerating expansion.  Only 
SNAP, with its systematics bounded measurements and broad redshift 
reach, looks promising for 
uncovering this clue.  Until recently this variation was parametrized 
through a linear expansion in redshift 
\beq 
w(z)=w_0+w_1z,
\label{wz}\eeq 
and SNAP measurements together with a prior on $\om$ of 0.03 
could determine $w_1$ to $\pm0.3$. 

However as we probe the expansion history to redshifts $z>1$ we would 
like a more accurate representation of the field evolution than 
Eq.~\ref{wz}.  Indeed if we wish to combine complementary cosmological 
probes, such as CMB data with supernovae, then clearly such a 
simple parametrization is insufficient.  Linder \cite{li02} has 
developed a new parametrization in terms of expansion factor $a$ 
that removes these difficulties and 
greatly improves the accuracy of reconstruction of the field behavior 
and the cosmological parameter estimation, 
\beq 
w(a)=w_0+w_a(1-a), 
\label{wa}\eeq 
while retaining simplicity and a two parameter phase space. 

\begin{figure}[!htb]
\centerline{\epsfxsize=3.0in\epsfbox{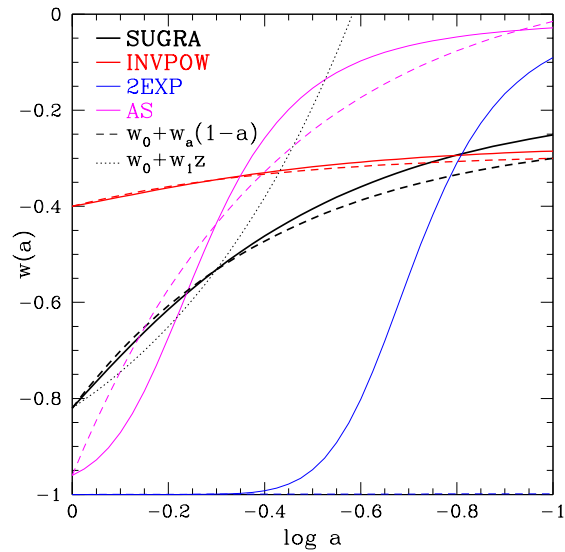}} 
\caption{$ $The 
equations of state of four dark energy potentials 
from 
${}^{11}$ 
are plotted as a function of expansion factor.  
Dashed lines show the reconstruction 
from the simple, new parametrization of Linder ${}^{10}$. 
The dotted 
line gives the old result for the SUGRA case. 
\label{wareal}} 
\end{figure}

Figure \ref{wareal} illustrates the success in modeling various 
dark energy models.  While the $w_1$ parametrization blows up for 
high redshifts, and indeed by $z=1.7$ is already off by 27\% for the 
supergravity inspired SUGRA model, the new parametrization stays 
well behaved and is within 
3\% of the true SUGRA equation of state.  More important observationally 
is the distance-redshift relation, and this parametrization recreates 
it for SUGRA to within 
0.2\% out to the CMB last scattering surface at $z=1100$ .

\begin{figure}[!htb]
\centerline{\epsfxsize=3.0in\epsfbox{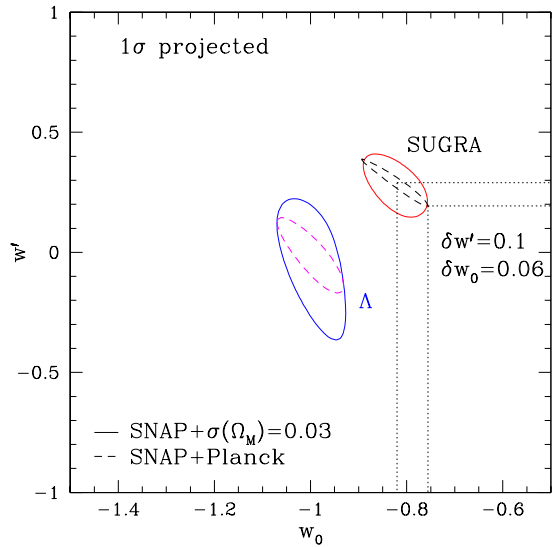}}   
\caption{Dark energy parameter estimation contours ($w'=w_a/2$) are 
shown for SNAP alone and SNAP in complementarity with Planck.  Note the 
estimation is sensitive to the region in phase space; here we plot 
results for the underlying model being either a cosmological constant 
or a supergravity inspired model.  Detection of time variation 
in the dark energy equation of state is possible in the SUGRA case 
at 99\% confidence. 
\label{lamsug}}
\end{figure}

The well behaved nature of this parametrization allows CMB data to be 
conjoined naturally with the supernovae measurements.  This has great 
advantages due to their complementarity \cite{fhlt}.  For example, 
SUGRA predicts $w_a=0.58$ and one might consider 
$w'\equiv dw/d\ln(1+z)|_{z=1}=w_a/2$ a natural measure of 
time variation.  It is 
directly related to the scalar field potential slow roll factor $V'/V$,  
and $z\sim 1$ is the region where the scalar field is most likely to be 
evolving as the 
epoch of matter domination begins to change over to dark energy 
domination.  SNAP plus data from the Planck CMB mission could 
estimate this important quantity to 
$\sigma(w')\approx0.1$, i.e.~this could demonstrate 
time variation of the dark energy equation of state at the $\sim$99\% 
confidence level (cf.~Figure \ref{lamsug}).  

\section{Summary} 

Dark energy is a crucial puzzle for 
fundamental physics, cosmology, and gravitation.  But upcoming 
experiments, especially the proposed SNAP satellite, can probe 
its nature by constructing datasets both precise and bounded in 
systematic effects.  Adding Planck CMB data, we have a 
real hope for exploring the time variation of the equation 
of state, a critical guide to the underlying physics.  Furthermore, 
mapping the recent expansion history of the universe can test the 
cosmological framework \cite{li02}, including higher dimensions, 
leading to new discoveries. 

\section*{Acknowledgments}

This work was 
supported in part by the Lawrence Berkeley National Laboratory of the 
U.S.~Department of Energy under contract DE-AC03-76SF0098.  I thank 
Neil Spooner for inviting me to participate in such a stimulating conference.


\begin{thebibliography}{99}

\bibitem{perl} 
        S.~Perlmutter {\it et al.},  {\it Ap.~J.} {\bf 517}, 565 (1999). 

\bibitem{hiz}
        A.~Riess {\it et al.}, {\it Astron.~J.} {\bf 116}, 1009 (1998). 

\bibitem{knox} 
 L.~Knox, N.~Christensen, and C.~Skordis, {\it Ap.~J.~Lett.} {\bf 563}, 
        L95 (2001), astro-ph/0109232. 

\bibitem{bond}
	J.R.~Bond {\it et al.}, astro-ph/0210007. 

\bibitem{teghamxu}
	M.~Tegmark, A.J.S.~Hamilton, and Y.~Xu, to appear in {\it MNRAS}, 
	astro-ph/0111575. 

\bibitem{efst}
	G.~Efstathiou {\it et al.}, astro-ph/0109152. 

\bibitem{snap}
	http://snap.lbl.gov; 
	G.~Aldering {\it et al.}, to appear in the {\it Proceeding of SPIE} 
  {\bf 4835}, astro-ph/0209550; 
	E.V.~Linder, to appear in the proceedings of the UCLA conference 
DM2002. 

\bibitem{lh02}
	E.V.~Linder and D.~Huterer, astro-ph/0208138. 

\bibitem{coping} 
	D.~Branch, S.~Perlmutter, E.~Baron, and P.~Nugent, astro-ph/0109070. 

\bibitem{li02}
        E.V.~Linder, astro-ph/0208512. 

\bibitem{corcop}
        P.S.~Corasaniti and E.J.~Copeland, astro-ph/0205544.  

\bibitem{fhlt} 
        J.A.~Frieman, D.~Huterer, E.V.~Linder, and M.S.~Turner, 
        submitted to {\it Phys.~Rev.~D}, astro-ph/0208100.  


\end{thebibliography}
\end{document}